\documentclass[useAMS,usenatbib]{mn2e}

\usepackage{epsfig}

\usepackage{times}

\def\sax{{\it BeppoSAX}}

\title[On the GRB Variability-Peak Luminosity Correlation]{The GRB Variability/Peak
Luminosity Correlation: new results}
\author[C. Guidorzi et al.]{C. Guidorzi$^{1,2}$\thanks{E-mail:
crg@astro.livjm.ac.uk},
F. Frontera$^{2,3}$, E. Montanari$^{2,4}$, F. Rossi$^{2}$, L. Amati$^{3}$,
\newauthor A. Gomboc$^{1,5}$, K. Hurley$^{6}$, C.G. Mundell$^{1}$\\
$^{1}$Astrophysics Research Institute, Liverpool John Moores University,
Twelve Quays House, Birkenhead, CH41 1LD, UK\\
$^{2}$Dipartimento di Fisica, Universit\`a di Ferrara, via Paradiso 12,
44100 Ferrara, Italy\\
$^{3}$Istituto Astrofisica Spaziale e Fisica Cosmica, section of Bologna, CNR/INAF,
Via Gobetti 101, 40129 Bologna, Italy\\
$^{4}$ISA ``Venturi'', Modena, Italy\\
$^{5}$Faculty of Mathematics and Physics, University in Ljubljana,
Jadranska 19, 1000 Ljubljana, Slovenia\\
$^{6}$Space Sciences Laboratory, University of California at Berkeley, 7 Gauss Way,
Berkeley, CA 94720-7450, USA}
\begin{document}

\date{}

\pagerange{\pageref{firstpage}--\pageref{lastpage}} \pubyear{2002}

\maketitle

\label{firstpage}

\begin{abstract}

We report test results of the correlation between time variability
and peak luminosity of Gamma-Ray Bursts (GRBs), using a larger sample
(32) of GRBs with known redshift than that available to \citet{Reichart01}, and using
as variability measure that introduced by these authors. The results
are puzzling. Assuming an isotropic-equivalent peak luminosity, as done 
by \citet{Reichart01}, a correlation is still found, but it is less relevant,
and inconsistent with a power law as previously reported.
Assuming as peak luminosity that corrected for GRB beaming for a subset of 16 GRBs
with known beaming angle, the correlation becomes little less significant.
\end{abstract}


\section{Introduction}
Despite the small number of Gamma-Ray Bursts (GRBs) with known redshift
(several dozens), several correlations between
intrinsic temporal or spectral parameters of the GRB prompt
emission and GRB energetics have been discovered in the last
seven years. \citet{Norris00} found an anticorrelation
between peak luminosity and the spectral lag (obtained by
cross-correlating the time profiles of the same GRB in various
energy bands), according to which more luminous bursts 
exhibit shorter lags. \citet{Salmonson02}
discovered a positive correlation between the spectral lag of
the gamma-ray prompt emission and the jet-break time of the afterglow
decay, according to which a small break time corresponds to
a small lag and consequently to a high peak luminosity of the GRB.
Concerning the temporal properties of GRB time profiles, evidence
has been found for a positive correlation between temporal variability
of the light curves and isotropic-equivalent peak luminosity for
the GRBs with known redshift (Reichart et al. 2001, hereafter
R01; Fenimore \& Ramirez-Ruiz 2000). 

Moreover \citet{Reichart03} have shown that
the variability vs. peak luminosity correlation could also hold
true for X-Ray Flashes (XRFs; see \citet{Heise01}).
As a consequence of the mentioned correlations, a correlation between
time variability and spectral lag is also expected and
confirmed for a large sample of BATSE bursts \citep{Schaefer01}.
The variability vs. peak luminosity correlation has been explained by several authors
(e.g., \citet{Kobayashi02, Meszaros02}) mainly within the framework of the standard
fireball model, according to which internal shocks between ultra relativistic shells
are responsible for the pulse-like structure of the GRB prompt emission, 
while the smooth afterglow emission is due to external shocks between the
fireball wind and the matter surrounding the GRB progenitor
(e.g., \citet{Piran04} for a review).

GRB variability-connected properties are thought to be more
sensitive to the bulk Lorentz factor $\Gamma$ and, if the GRB emission
is beamed, to the jet opening angle and/or the viewing angle 
(e.g., \citet{Salmonson02,Ioka01}).
Within the fireball model, there are different mechanisms that
could account for different time variabilities, also giving possible
explanations for XRF properties \citep{Meszaros02}.
In addition, the ``cannonball model'' for GRBs \citep{Dado02} also seems to
explain the variability vs. peak luminosity correlation \citep{Plaga01}.

From the above correlations also luminosity estimators have tentatively
been derived \citep{Reichart01,Fenimore00,Schaefer01} to investigate
general properties of GRBs, such as the luminosity function and the possible
link with the star formation rate. In addition, empirical redshift
indicators have been proposed based on the calibration derived with the small
sample of GRBs with known redshift, making use of the X- and gamma-ray
observations alone \citep{Atteia03a,Bagoly03}.

In this work we test the variability vs. peak luminosity correlation
using the variability definition given by R01.
We used a sample of 32 GRBs with known redshift.
Furthermore we studied the same correlation by replacing the isotropic-equivalent
peak luminosity with that corrected for beaming for a subset of 16 GRBs with known
collimation angle provided by \citet{Ghirlanda04}.

In Section~\ref{s:obs}, we discuss our sample of GRBs; in Section~\ref{s:var}
we discuss the time variability analysis; in Section~\ref{s:lum} we estimate 
the peak luminosity of the GRB in our sample and compare it with the R01 results.
In Section~\ref{s:results} we present our results on variability/peak luminosity
correlation, in Section~\ref{s:disc} we discuss our results.

%%%%%%%%%%%%%%%%%%%%%%%%%%%%%%%%%%%%%%%%%%%%%%%%%%%%%%%%%
\section[]{The GRB sample}
\label{s:obs}
%%%%%%%%%%%%%%%%%%%%%%%%%%%%%%%%%%%%%%%%%%%%%%%%%%%%%%%%%

%---------------------------------------------------
\subsection[]{GRBs with known redshift}
\label{subs:obs:z}
%---------------------------------------------------
The sample of 32 GRBs with known redshift includes 16 GRBs detected
by the Gamma-Ray Burst Monitor (GRBM) \citep{Feroci97,Frontera97,Costa98}
on board the \sax\ satellite \citep{Boella97} during the period 1997--2002, two by
the BATSE experiment \citep{Paciesas99}  aboard the {\it Compton Gamma--Ray
Observatory} ({\it CGRO}), six by the FREGATE instrument aboard {\it HETE-II}
\citep{Atteia03b}, one by Konus/{\it WIND} \citep{Aptekar95}, one by {\it Ulysses} \citep{Hurley92}
and six by BAT/{\it Swift} \citep{Gehrels04}.
Eight of the 16 GRBs detected with the \sax\ GRBM were also detected with BATSE.
We used public archives for GRB data obtained with 
BATSE\footnote{ftp://cossc.gsfc.nasa.gov/compton/data/batse/ascii\_data/64ms/},
{\it HETE-II}\footnote{http://space.mit.edu/HETE/Bursts/Data/},
Konus/{\it WIND}\footnote{http://lheawww.gsfc.nasa.gov/docs/gamcosray/legr/\\bacodine/konus\_grbs.html},
and BAT/{\it Swift}\footnote{http://swift.gsfc.nasa.gov/}.
Table~\ref{tab:total_z} reports the list of the GRBs in our sample with mentioned
the spacecraft that detected it.
When the same GRB has been detected by more than one instrument,
we first checked the consistency of the results derived from different data
sets and then concentrated on the instrument which had the best SNR.

The time binning of the GRB light curves in our sample, which was used to derive
the time variability, was the following: 7.8125~ms for the GRBM data, 
64~ms for BATSE, 164~ms for {\it HETE-II}, 64~ms for Konus/{\it WIND}, 31.25~ms for {\it Ulysses}. 
In the case of BAT/{\it Swift} we made use of the event files and extracted
the mask-tagged light curves with a binning time of 8~ms.
Given that for the GRBs detected with the \sax\ GRBM, the high-resolution 
(7.8125-ms binning) time profiles are available in the 40--700~keV
energy band, for the others we used the light curves in the energy bands
which have the largest intersection with the 40--700~keV band: 110--320~keV
(channel 3) for BATSE, 30--400~keV (band C) for FREGATE, 25--100~keV for {\it Ulysses}
and 50--200~keV for Konus/{\it WIND}.
In order to match the GRBM band, for BAT/{\it Swift} we extracted the light curves
from the event files in the 40--350~keV band.

For GRB990510, given that the 7.8125-ms GRBM light curve is not available, 
we preferred to use 64-ms BATSE data rather than the 1-s GRBM light curve. 

Six GRBs (980613, 011211, 021004, 050126, 050318 and 050416) with known redshift were
not included in our sample due to their low signal, which prevented us from deriving a 
statistically significant variability estimate. Another GRB (021211, detected with
{\it HETE-II}) was not included in the sample, due to the high ratio between binning time
and smoothing time which could bias the variability estimate (see Section~\ref{s:var}).
Other GRBs with known redshift (000301C, 000418, 000926) detected by both
Konus/{\it WIND} and {\it Ulysses} were not included in our sample, because unfortunately 
both Konus public and {\it Ulysses} data cover their light curves only partially.

In the case of BATSE (970828 and 000131) the usual 4-channel 64-ms light curves
are not available. Thus we made use of the 16-channel MER spectra acquired along
either entire GRB with an integration time of 64~ms.
Therefore we rebinned the 16-energy-channel MER data both spectrally and temporally in
order to reproduce as much as possible the 4-channel scheme of BATSE 64-ms light profiles.
As we discuss below, we relied on coarse time resolution light curves only when the
overall duration of the GRB was very long compared to the binning time.

In general the data available cover the entire time profile of the GRBs
in our sample. However there are some exceptions. In the case of the GRBM events,
given that the high-resolution data cover 8~s before the trigger 
time and 98~s after it, in the case of the longest events 
(990506 and 010222, $T_{90}=129$~s and $T_{90}=97$~s, respectively), it is not true.
In these cases, the measure of time variability was obtained summing 
the variability in the part covered by the 7.8125-ms bins with that  
in the part covered by 1-s ratemeters (the tail of the burst).

GRB000210 ($T_{90} = 8.1$~s) suffers from a 2.5-s long gap
due to corrupted high-resolution data occurred in the middle of the burst profile.
Using the 1-s data, the mean 7.8125-ms light profile in the gap was reconstructed
adding to the mean profile a Poisson noise. The value of GRB time 
variability so derived was found not to significantly change, even adding in the gap
a non-Poisson noise compatible with the 1-s time profile.

For each GRB detected with the GRBM, we considered the light curves of the 
two most illuminated units and checking whether the best signal-to-noise ratio (SNR) 
was obtained from a single unit or by summing the two units.

We found that for the 11 GRBs detected by GRBM and Wide Field Cameras (WFCs)
\citep{Jager97}, the best signal is obtained from a single 
GRBM unit (that with the larger area exposed to the GRB).
For the five bursts detected with the GRBM but not with the WFCs the best
signal was obtained by summing the two most illuminated units:
units 1 and 4 (980703), 3 and 4 (990506, 020405), 2 and 4 (991216, 010921).
In principle the operation of adding the counts of different units is questionable
because of dead time, as it will be discussed in Section~\ref{s:var}.
In practice, for the above cases we made sure that the results were consistent with
those obtained when considering only the most illuminated unit for each GRB.
This has been found to be no longer true, i.e. the correction for dead time becomes
not negligible, when considering very small smoothing time-scales (Rossi et~al.,
in preparation).
% In the case of BATSE GRBs, we used the most illuminated unit because of the
% better SNR, combined with the larger number of detection units: eight
% Large Area Detectors (LADs) instead of the four GRBM units.

\begin{table*}
 \centering
 \begin{minipage}{150mm}
  \caption{Variability vs. Peak Luminosity for 32 GRBs with known redshift (1$\sigma$ errors).}
  \label{tab:total_z}
  \begin{tabular}{lllrrrr}
\hline
%\noalign{\smallskip}
GRB &  $z$ & Mission$^{\rm (a)}$ & $T_{f=0.45}$ & $V_{f=0.45}$ & Peak Lum. $L^{\rm (b)}$ & $z$ Refs.$^{\rm (c)}$\\
Name & Redshift &  & (s)        &   & ($10^{50}$~erg s$^{-1}$) & \\
%\noalign{\smallskip}
\hline
%\noalign{\smallskip}
970228 & 0.695 & BS/U/K       & $   2.2$ & $0.223_{-0.017}^{+0.018}$   & $  48.7 \pm   9.9$ & 1\\
970508 & 0.835 & B/BS/U/K     & $   2.4$ & $0.023_{-0.013}^{+0.013}$   & $   9.43 \pm   1.89$ & 2\\
%970828 & 0.958 & B/U/K/S      & $ 31.10$ & $0.0397_{-0.0003}^{+0.0003}$& $ 120.0 \pm  40.0$ & 3\\
970828 & 0.958 & B/U/K/S      & $  12.9$ & $0.101_{-0.002}^{+0.002}$   & $ 120.0 \pm  40.0$ & 3\\
971214 & 3.418 & BS/B/U/K/N/R & $   4.4$ & $0.110_{-0.012}^{+0.012}$   & $ 360. \pm  65.$ & 4\\
980425 & 0.0085& B/BS/U/K     & $   4.7$ & $0.049_{-0.048}^{+0.048}$   & $0.0007 \pm 0.0002$ & 5\\
980703 & 0.966 & BS/B/U/K/R   & $   3.2$ & $0.044_{-0.007}^{+0.007}$   & $  26.4 \pm   5.6$ & 6\\
990123 & 1.6   & BS/B/U/K     & $  12.8$ & $0.112_{-0.002}^{+0.002}$   & $ 840. \pm 121.$ & 7\\
990506 & 1.3   & BS/B/U/K/R   & $   8.6$ & $0.270_{-0.005}^{+0.005}$   & $ 583. \pm 121.$ & 8\\
990510 & 1.619 & B/BS/U/K/N   & $   3.2$ & $0.214_{-0.008}^{+0.005}$   & $ 300. \pm  50.$ & 9\\
990705 & 0.86  & BS/U/K/N     & $   8.0$ & $0.178_{-0.003}^{+0.003}$   & $ 134. \pm  21.$ & 10,11\\
990712 & 0.434 & BS/U/K       & $   4.1$ & $0.042_{-0.017}^{+0.017}$   & $   5.4 \pm   1.0$ & 12\\
991208 & 0.706 & K/U/N        & $   5.1$ & $0.082_{-0.003}^{+0.003}$   & $ 290. \pm 100.$ & 13\\
991216 & 1.02  & BS/B/U/N     & $   2.6$ & $0.193_{-0.002}^{+0.002}$   & $1398. \pm 200.$ & 14\\
%000131 & 4.5   & B/U/K/N      & $   8.8$ & $0.170_{-0.002}^{+0.004}$   & $3600. \pm 900.$ & 15\\
000131 & 4.5   & B/U/K/N      & $   8.0$ & $0.187_{-0.005}^{+0.005}$   & $3600. \pm 900.$ & 15\\
000210 & 0.846 & BS/U/K       & $  1.59$ & $0.026_{-0.002}^{+0.002}$   & $480. \pm  50.$  & 16\\
000911 & 1.058 & U/K/N        & $   5.2$ & $0.077_{-0.034}^{+0.034}$   & $ 360. \pm  60.$ & 17\\
010222 & 1.477 & BS/U/K       & $  6.62$ & $0.201_{-0.003}^{+0.003}$   & $ 801. \pm 119.$ & 18\\
010921 & 0.45  & BS/H/U/K     & $   5.3$ & $0.038_{-0.016}^{+0.016}$   & $   8.0 \pm   2.0$ & 19\\
011121 & 0.36  & BS/U/K/O     & $   8.3$ & $0.049_{-0.002}^{+0.002}$   & $  19.9 \pm   3.1$ & 20\\
020124 & 3.198 & H/U/K        & $   8.8$ & $0.203_{-0.032}^{+0.031}$   & $300. \pm 60.$ & 21\\
020405 & 0.69  & BS/U/K/O     & $   9.9$ & $0.168_{-0.007}^{+0.007}$   & $  71.4 \pm  11.2$ & 22\\
020813 & 1.25  & H/U/K/O      & $  17.4$ & $0.248_{-0.007}^{+0.007}$   & $ 340. \pm  70.$ &23\\
030226 & 1.98  & H/K/O        & $  26.6$ & $0.042_{-0.015}^{+0.015}$   & $  25.0 \pm   5.0$ &24\\
030328 & 1.52  & H/U/K        & $  24.9$ & $0.051_{-0.005}^{+0.005}$   & $  90. \pm  18.$ &25\\
030329 & 0.168 & H/U/K/O/RH   & $   4.9$ & $0.105_{-0.007}^{+0.007}$   & $   6.1 \pm 1.2$ & 26\\
041006 & 0.712 & H/K/RH       & $   8.0$ & $0.052_{-0.002}^{+0.002}$   & $   66. \pm 10.$ & 27\\
050315 & 1.949 & BSw          & $  12.3$ & $0.042_{-0.031}^{+0.032}$   & $   38. \pm 8.$ & 28\\
050319 & 3.24  & BSw          & $   3.6$ & $0.061_{-0.030}^{+0.032}$   & $   84. \pm 20.$ & 29\\
050401 & 2.90  & BSw          & $   4.4$ & $0.195_{-0.029}^{+0.028}$   & $  740. \pm 100.$ & 30\\
050505 & 4.27  & BSw          & $   9.0$ & $0.205_{-0.044}^{+0.043}$   & $  250. \pm 50.$ & 31\\
050525 & 0.606 & BSw          & $   2.0$ & $0.111_{-0.003}^{+0.003}$   & $   80. \pm 10.$ & 32\\
050603 & 2.821 & BSw          & $   1.2$ & $0.245_{-0.034}^{+0.037}$   & $ 1200. \pm 200.$ & 33\\
\hline
\end{tabular}
\begin{list}{}{}
\item[$^{\rm (a)}$]Mission: BS (\sax), B (BATSE/{\it CGRO}), K (Konus/{\it WIND}), H ({\it HETE-II}), 
U ({\it Ulysses}), S ({\it SROSS-C}), N ({\it NEAR}), R ({\it RossiXTE}), O ({\it Mars Odyssey}),
RH ({\it RHESSI}), BSw (BAT/{\it Swift}): the data used are taken from the first mission mentioned.
\item[$^{\rm (b)}$]Isotropic-equivalent peak luminosity in $10^{50}$~erg s$^{-1}$
  in the rest-frame 100--1000~keV band, for peak fluxes measured on a 1-s time-scale,
  $H_0 = 65$ km s$^{-1}$ Mpc$^{-1}$, $\Omega_m = 0.3$, and $\Omega_{\Lambda} = 0.7$.
\item[$^{\rm (c)}$]References for the redshift measurements:
(1) \citet{Djorgovski99}, (2) \citet{Metzger97}, (3) \citet{Djorgovski01a}, (4) \citet{Kulkarni98}, (5) \citet{Tinney98},
(6) \citet{Djorgovski98}, (7) \citet{Kulkarni99}, (8) \citet{Bloom03}, (9) \citet{Beuermann99}, 
(10) \citet{Amati00}, (11) \citet{Lefloch02}, (12) \citet{Galama99}, (13) \citet{Dodonov99}, (14) \citet{Vreeswijk99},
(15) \citet{Andersen00}, (16) \citet{Piro02}, (17) \citet{Price02a}, (18) \citet{Garnavich01},
(19) \citet{Djorgovski01b}, (20) \citet{Infante01}, (21) \citet{Hjorth03}, (22) \citet{Masetti02},
(23) \citet{Price02b}, (24) \citet{Greiner03a}, (25) \citet{Martini03}, (26) \citet{Greiner03b},
(27) \citet{Fugazza04}, (28) \citet{Kelson05}, (29) \citet{Fynbo05a}, (30) \citet{Fynbo05b},
(31) \citet{Berger05_a}, (32) \citet{Foley05}, (33) \citet{Berger05_b}.
\end{list}
\end{minipage}
\end{table*}

As far as the 8 bursts detected with both GRBM and BATSE (970508, 971214, 980425, 980703,
990123, 990506, 990510, 991216) are concerned, we used the \sax\ data for
971214, 980703, 990123, 990506, and 991216, for which the higher time resolution of the GRBM turned
out to be essential for a better variability estimate, while for the remaining 3 GRBs
(970508, 980425, and 990510) we used the BATSE data given the better SNR,
after verifying the mutual consistency of the GRBM and BATSE variability results.

%%%%%%%%%%%%%%%%%%%%%%%%%%%%%%%%%%%%%%%%%%%%%%%%%%%%%%%%%
\section{Variability measure}
\label{s:var}
%%%%%%%%%%%%%%%%%%%%%%%%%%%%%%%%%%%%%%%%%%%%%%%%%%%%%%%%%
We adopted the variability measure given by R01, slightly modified for two 
corrections which could affect the result:  the instrument dead time,
and a small non-Poisson noise present in the GRBM background data.
The variability measure used by R01 was defined as a properly normalised
mean square deviation of the intrinsic light curve of a GRB in a given
energy band from a smoothed one. For a discrete light curve made of
N bins, the variability measure, according to R01, is given by:
\begin{eqnarray}
\label{eq:varpoiss1}
~~~~~~~V_{f,P} & = & \frac{\sum_{i=1}^N [S_i(C_j,N_z) -
S_i(C_j,N_f)]^2}{\sum_{i=1}^N [S_i(C_j,N_z) - B_i]^2},
\end{eqnarray}
where as intrinsic light curve we mean the GRB light curve in the source-frame,
$N_f$ is the number of data bins corresponding to the smoothing time scale $T_f$ 
defined by R01 as the shortest cumulative
time interval during which a fraction $f$ of the total counts above 
background has been collected, $C_j$ and $B_i$ are the original GRB
(source plus background) and background counts in the bins $j$ and $i$, respectively,
in the observer frame, the index
$P$ means that the variability measure is inclusive of the Poisson noise.
$S_i(C_j,N_x)$ is roughly the mean counts on $N_x$ bins ($x= z$ or $f$) 
centred around the $i$-th bin:
\begin{eqnarray}
\label{eq:smear}
S_i(C_j,N_x) = \frac{1}{N_x}\big[\sum_{j=i-n_x}^{i+n_x}C_j +
~~~~~~~~~~~~~~~~~~~~~~~\nonumber\\
~~~~~~~~~~~~~~~~~~~~~~+\big(\frac{N_x-1}{2}-n_x\big) \ (C_{i-n_x-1}+C_{i+n_x+1})\big].
\end{eqnarray}
$N_z$ is the number of bins in the observer-frame, which corresponds
to 1 bin in the source-frame. Assuming as time duration of 1 bin in the source-frame
the shortest binning $\Delta t$ of the data (e.g., in the case of the \sax\ GRBM,
$\Delta t = 7.8125$~ms), in the observer-frame the number of bins, depending on the GRB  
redshift $z$, for relativistic time dilation and narrowing of the light 
curves at high energies \citep{Fenimore95}, is given by $N_z = (1+z)^\beta$
with $\beta \approx 0.6$. Thus $N_z$ can take values other than integers and  
$n_x$ is the truncated integer value of $(N_x-1)/2$.
%
% Unlike R01, who, using only BATSE data, had the same binning for all GRBs in their sample, 
% we have data from different instruments and $\Delta t$ is different for each of them
% (see Section~\ref{s:obs}). 
% NOTE (CRISTIANO): NOT REALLY TRUE: R01 USED DATA FROM DIFFERENT INSTRUMENTS TOO
% (KONUS, E.G.)
%

R01 found that the best luminosity estimator is obtained when using $f=0.45$;
for this reason, we fixed $f=0.45$.

The variability $V_{f,P}$ can also be written as follows:
\begin{eqnarray}
~~~~~~~V_{x,P} & = & \frac{\sum_{i=1}^N (\sum_{j=1}^Na_{ij}C_j)^2}{\sum_{i=1}^N (\sum_{j=1}^Nb_{ij}C_j-B_i)^2}
\label{eq:varpoiss2}
\end{eqnarray}
where the coefficients $a_{ij}$ and $b_{ij}$, for each GRB,  are computed by 
comparing eq.~\ref{eq:varpoiss2} with eq.~\ref{eq:varpoiss1} through eq.~\ref{eq:smear}.

Following R01, after subtraction of the Poisson variance the variability measure
is given by 
\begin{equation}
  V_f = \frac{\sum_{i=1}^N 
    [(\sum_{j=1}^Na_{ij}C_j)^2-\sum_{j=1}^Na_{ij}^2\,C_j]}{\sum_{i=1}^N 
    [(\sum_{j=1}^Nb_{ij}C_j-B_i)^2-\sum_{j=1}^Nb_{ij}^2\,C_j]}
  \label{eq:var_r01}
\end{equation}
which is the expression used by R01 to evaluate the variability of the GRBs in their
sample.
We slightly modified the above expression by taking also into account the dead time,
which is known to affect the Poisson variance of a stationary process
\citep{Mueller73,Mueller74,Libert78}. In the case of a stationary Poisson 
process with measured mean rate $\mu$,
the variance of its counts in the time bin $\Delta t$, which is given by $\mu\Delta t$
in absence of dead time, becomes  $\mu\Delta t\ (1-\mu\tau)^2$ in the asymptotic 
limit $\tau/\Delta t\ll 1$, where $\tau$ is the dead time. In the case
of the \sax\ GRBM $\tau=4\mu$s, $\tau/\Delta t\simeq5\times10^{-4}$ for the 
shortest bin duration $\Delta t=7.8125$~ms. In the same limit $\tau/\Delta t\ll 1$,
the same correction factor $(1-\mu\tau)^2$ applies to the white noise level of 
the power spectral density (PSD) estimate \citep{Frontera78,Klis89}.
It is shown \citep{Frontera79} that the same correction factor holds when 
the process is non-stationary, like GRBs or flares.
Potentially our variability calculations could be sensitive to dead time, especially for
those GRBs with huge peak count rates, like in the case of 990123 ($\sim$16,000 cts/s with
GRBM), for which, around the peak, the true variance is $\sim0.9$ times the 
measured counts. % (see Fig.~\ref{f:deadtime_variance}).

In addition, we corrected  for a slight (a few percent) non-Poisson noise 
found in the GRBM high-resolution data. This noise increases the Poisson variance
by a factor $r_{np}$ which ranges from 1.027 to 1.049, depending on the
detection unit, for the GRBM data after November 1996 \footnote{During the 
first months of \sax\ operation the non-Poisson noise of the GRBM
was much higher, due to the too low energy threshold (around 20 keV)
set at the beginning of the mission \citep{Feroci97}.}.

Taking into account both dead time and non-Poisson noise, the right terms
to be subtracted in the numerator and denominator of eq.~\ref{eq:varpoiss2}
become $\sum_{j=1}^Na_{ij}^2C_j\ r_j$ and $\sum_{j=1}^Nb_{ij}^2C_j\ r_j$, 
respectively (see, for comparison, eq.~\ref{eq:var_r01}), where
\begin{equation}
  r_j = r_{np}\ \big(1-C_j\,\frac{\tau}{\Delta\,t}\big)^2
\end{equation}

As a consequence, the expression we used to estimate the net GRB time
variability is given by:
\begin{equation}
  V_f = \frac{\sum_{i=1}^N 
    [(\sum_{j=1}^Na_{ij}C_j)^2-\sum_{j=1}^Na_{ij}^2\,C_j\,r_j]}{\sum_{i=1}^N 
    [(\sum_{j=1}^Nb_{ij}C_j-B_i)^2-\sum_{j=1}^Nb_{ij}^2\,C_j\,r_j]}
  \label{eq:var}
\end{equation}
We used as statistical uncertainty  $\sigma_{V_f}$ on the variability measure
that given by R01 (eq.~8) properly
modified to take into account the correction factor $r_j$.

We found that the variability measure is not sensitive to dead time
corrections for long GRBs, in which $T_{f=0.45}$ is much longer than the
bin time, while it is significantly modified for relatively short GRBs
exhibiting sharp intense pulses. 

%----------------------------------------------------------------------------
\subsection{Variability dependence on binning time}
\label{ss:bintime}
%----------------------------------------------------------------------------

In order to understand how time binning affects the GRB variability, 
for the brightest GRBs, detected with either GRBM or BATSE we evaluated the
variability measure (eq.~\ref{eq:var}) as a function of the binning time of the
data. The result is that the variability is better estimated for very short 
time durations of the data bins with respect to the smoothing 
time-scale $T_s= T_{f=0.45}$. More specifically, it results that the variability 
significantly decreases for few~0.01$<\Delta\,t/T_s<$~few~0.1, and becomes
unreliable when this ratio becomes still higher, i.e. for $\Delta\,t/T_s>$few~0.1.
%
%Figure 1a
%
\begin{figure}
\begin{center}
\centerline{\includegraphics[width=9.0cm]{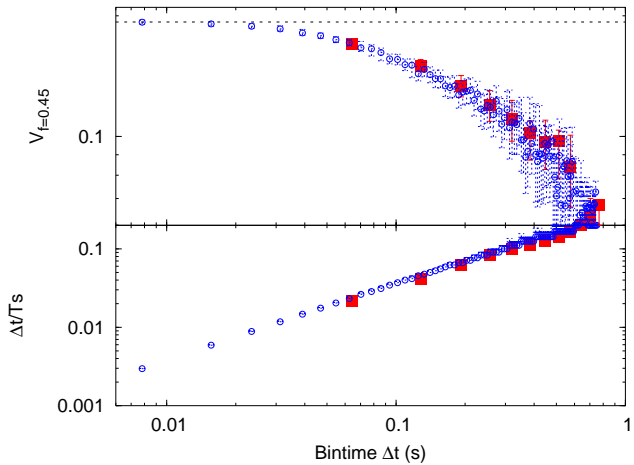}}
\caption{Top panel: Variability of 991216 as a function of binning time $\Delta t$
using both GRBM (blue circles) and BATSE (red squares) data sets.
Black dotted line represents the asymptotic value of
$V_{f=0.45}$ as derived with GRBM data using small binning times.
Also shown is the ratio between binning time and smoothing time-scale (bottom panel).}
\label{f:991216_var_bintime}
\end{center}
\end{figure}

On the other side, the bin time should be long enough to collect a good number of 
photons (typically at least 20 per bin on average) to ensure the Gaussian limit
and take over the effects of statistical fluctuations.
Thus we rejected those GRBs whose data sets do not match the above requirements.
%
%Figure 1b
%
\begin{figure}
\begin{center}
\centerline{\includegraphics[width=9.0cm]{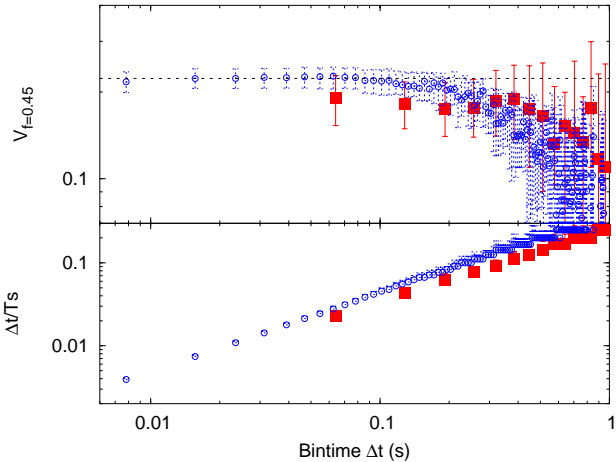}}
\caption{Top panel: Variability of 970228 as a function of binning time
$\Delta t$ using both GRBM (blue circles) and KONUS (red squares) data sets.
Black dotted line represents the asymptotic value of
$V_{f=0.45}$ as derived with GRBM data using small binning times.
Also shown is the ratio between binning time and smoothing time-scale (bottom panel).}
\label{f:970228_var_bintime}
\end{center}
\end{figure}

Figures~\ref{f:991216_var_bintime} and~\ref{f:970228_var_bintime} show
the illustrative cases of 991216 and 970228, respectively.
We calculated $V_{f=0.45}$ using both GRBM and BATSE (KONUS) data sets as a
function of the binning time for 991216 (970228).
For both GRBs, the variability seems to approach an asymptotic value for
decreasing values of binning time.
In the case of 991216, it appears that the original binning
time of BATSE, 64 ms, is a little too coarse since its correspondent value
of $V_{f=0.45}$ is significantly lower: we assume as asymptotic
value the measure obtained with the smallest binning time of GRBM data
and get $V_{f=0.45}=0.193\pm 0.002$, while the BATSE measure is
$0.170\pm 0.003$, i.e. $\sim$6-$\sigma$ apart.
Differently, in the case of 970228 the KONUS measure with the smallest
binning time of 64~ms yields a measure of $V_{f=0.45}$ which is apparently
consistent with the GRBM one (Fig.~\ref{f:970228_var_bintime}).
In the case of 991216 it is worth noting that the measure of $V_{f=0.45}$
turns out to be significantly underestimated with respect to the asymptotic
value as far as we assume binning times at least a few $10^{-2}$ times as high
as the smoothing time-scale.

In general, we noticed that for all the GRBs for which $V_{f=0.45}$ approaches
an asymptotic value for small binning times, different measures of $V_{f=0.45}$ are
still consistent with that, provided that the ratio between binning time and smoothing
time-scale is not too high ($\Delta t/T_s<$~few~$10^{-2}$).

Our final set of variability measures include only those GRBs for which
the three following requirements have been fulfilled with a single binning time:
1) smallness of ratio $\Delta t/T_s$, 2) asymptotic behaviour of $V_{f=0.45}$
as a function of binning time $\Delta t$ for small $\Delta t$, 3) Gaussian limit
of at least 20 counts per bin on average.

Following this guideline, we discarded the {\it HETE-II} bursts 021211 and 050408,
for which $\Delta\,t/T_s$ is around 0.2 and 0.08, respectively.
In the case of the couple of GRBs above considered, we infer that GRBM data
turned out to be essential in estimating the variability of 991216, since
BATSE data alone, although consistent with GRBM data for comparable binning times,
do not seem to approach an asymptotic value of $V_{f=0.45}$, while GRBM data do.
On the other side, in the case of 970228 KONUS data exhibit an asymptotic trend
towards small binning times; together with the fulfillment of the other two
requirements, KONUS time resolution is acceptable and yields a variability measure
which is consistent with the GRBM within errors.

%%%%%%%%%%%%%%%%%%%%%%%%%%%%%%%%%%%%%%%%%%%%%%%%%%%%%%%%%
\section{Peak Luminosity Estimate}
\label{s:lum}
%%%%%%%%%%%%%%%%%%%%%%%%%%%%%%%%%%%%%%%%%%%%%%%%%%%%%%%%%
The GRBs peak luminosities were estimated using the definition 
of luminosity distance in the source-frame 100--1000~keV energy band:
\begin{equation}
\displaystyle L \ =\  4 \pi D_L^2(z)\ \int_{100/(1+z)}^{1000/(1+z)} E \Phi(E)\,dE
\label{eq:lum}
\end{equation}
where $\Phi(E)$ is the measured spectrum (ph cm$^{-2}$s$^{-1}$keV$^{-1}$)
around the peak time, $D_L(z)$ is the
luminosity distance at redshift $z$, $E$ is energy expressed in keV.
By replacing $E'= E(1+z)$ we get:
\begin{equation}
\displaystyle L \ = \ \frac{4 \pi D_L^2(z)}{(1+z)^2}\ \int_{100}^{1000}{E'
\Phi\Big(\frac{E'}{1+z}\Big)\,dE'}
\label{eq:lum2}
\end{equation}
Formally, eq.~\ref{eq:lum2} is the same as eq.~9 of R01: there $D(z)$ is
the co-moving distance, which is equal to $D(z)=D_L/(1+z)$ if we consider
a flat Universe.
However, unlike R01 who used as $\Phi(E)$ the best-fitting Band model
\citep{Band93} to the average GRB count spectrum normalised to the peak 
count rate, we used for the GRBM data the best-fitting power-law spectrum 
($\Phi (E) = NE^{-\alpha}$) to the GRB peak count rate spectrum obtained 
from the 1-s ratemeters available in two channels (40--700 keV and $>100$~keV). 
When the 225-channel time-averaged spectrum was not
available, we added a conservative 10\% systematics to the peak luminosity uncertainties.
Thus, for the GRBM bursts, eq.~\ref{eq:lum2} becomes:
\begin{equation}
L \ = \ 4 \pi D_L^2(z)\ (1+z)^{\alpha -2}\,F_p
\label{eq:lum3}
\end{equation}
where $F_p=\int_{100}^{1000} NE'^{1-\alpha} dE'$ is the 100--1000~keV peak flux
measured in the observer frame (erg cm$^{-2}$s$^{-1}$).
%
%Figure 2
%
\begin{figure}
\includegraphics[width=10cm]{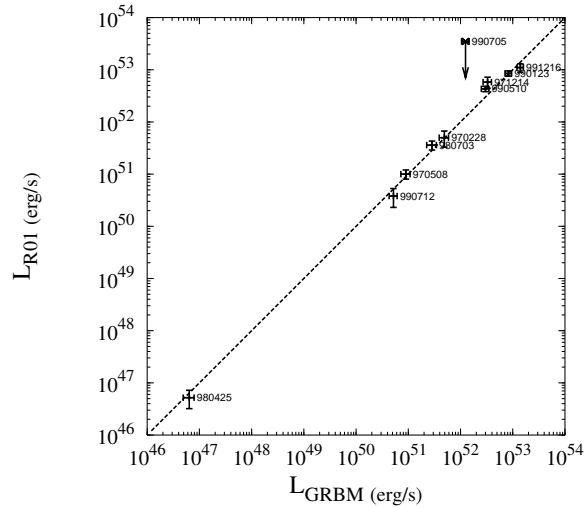}
\caption{Peak luminosity $L_{\textrm{GRBM}}$ derived with GRBM data
vs. peak luminosity $L_{\textrm{R01}}$ published by Reichart et al. using data
from BATSE, Konus/{\it WIND} and {\it Ulysses}, for a common sample of GRBs.
The dashed line shows the equation $L_{\textrm{R01}} = L_{\textrm{GRBM}}$. See text.}
\label{f:lum_unsrer_vs_reichart}
\end{figure}
In the case of GRBs with sharp peaks of $<1$~s duration (e.g. GRB000214), 
the peak luminosity obtained from 1-s ratemeters was further corrected by 
the ratio between the actual peak value and that derived from 1-s ratemeters.

For the GRBs in our sample not detected with the GRBM, we
used  the best-fitting parameters of $\Phi(E)$ available from the literature.
The best-fitting spectral parameters for {\it HETE-II} bursts were taken
from \citet{Sakamoto04}, except for the recent 041006 for which we used the
best-fitting cutoff power-law parameters published by {\it HETE-II} team on the
HETE web page ($E_{\rm 0}$=100.2~keV, $\alpha=1.367$).
For the {\it Ulysses} GRB000911 we made use of the best-fitting
parameters published by \citet{Price02a}, while for the Konus burst 991208
we used the parameter values given by R01.
For the BATSE GRB000131 we fitted the peak energy spectrum from MER
data in the range 30--1000~keV with the Band function
($\alpha=-0.56$, $\beta=-2.17$, $E_0=153$~keV, $\chi^2/{\rm dof} = 1.0$).
Likewise, for the BATSE GRB970828 the peak energy spectrum was fitted
with the Band function
($\alpha=-0.65$, $\beta=-2.56$, $E_0=269$~keV, $\chi^2/{\rm dof} = 1.1$).
For BAT/{\it Swift} we extracted from the event file the 1-s 80-channel
spectrum around the peak; for all the 6 BAT/{\it Swift} GRBs considered, the peak
spectrum was fitted with a simple power law in the 15-350~keV range, apart
from a couple of them (050525 and 050603) for which only the cut-off power-law
model yields a good fit. We then used eq.~\ref{eq:lum} to evaluate the
peak luminosity.

Our peak luminosity estimates are reported in Tables~\ref{tab:total_z}.

For the common sample of GRBs, our estimates of the peak luminosity are
fully consistent with those obtained by R01 (see Fig.~\ref{f:lum_unsrer_vs_reichart}).

%%%%%%%%%%%%%%%%%%%%%%%%%%%%%%%%%%%%%%%%%%%%%%%%%%%%%%%%%
\section{Results}
\label{s:results}
%%%%%%%%%%%%%%%%%%%%%%%%%%%%%%%%%%%%%%%%%%%%%%%%%%%%%%%%%

%--------------------------------------------------------
\subsection{GRBs with known redshift}
%--------------------------------------------------------
First of all we evaluated the time variability of the GRBs (13) common to our sample 
and to that by R01, in order to test the mutual consistency of our results 
with those obtained by R01.

\subsubsection{Variability}

\begin{figure}
\includegraphics[width=10cm]{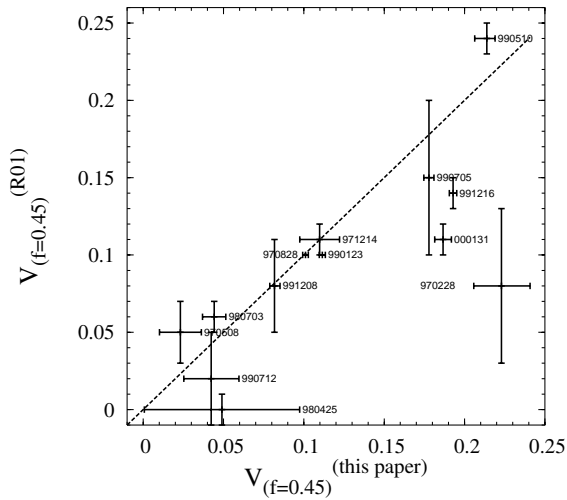}
\caption{Variability according to R01's definition derived in this paper
vs. Variability published by R01 for a common sample of 13 GRBs.
The dashed line shows the equation corresponding to equality. See text.}
\label{f:unsrer_vs_reichart}
\end{figure}

Figure~\ref{f:unsrer_vs_reichart} compares the two time variability estimates. As
can be seen, the results are well consistent with each other, except for three cases
(970228, 991216, and 000131).
%
%Figure 3
%

For each of these GRBs we investigated the reason
of the discrepant measure of $V_{f=0.45}$ first of all by trying to reproduce the
results by R01 using KONUS data alone for 970228 and BATSE data alone for the other
two.

\subsubsection{GRB~970228}
In order to reproduce R01's results for 970228 we used the same data set, i.e. the
light curve by KONUS. The only difference is that we used public data that include
a single light curve in the 50--200~keV energy band, while R01 used three different
energy bands: 10--45~keV, 45--190~keV and 190--770~keV.
R01 report the smoothing time-scale for each of the three energy channels and only
the global variability measure derived from merging the three different measures 
according to the procedure described therein.
Our measure of the smoothing time-scale is $2.82\pm0.32$~s to be compared with
that obtained by R01 for the same energy channel, i.e. 2.891~s (no error is reported),
thus consistent. Our variability measure with KONUS data is $0.19\pm0.04$
to be compared with R01's one, $0.08\pm0.05$. The measure obtained with GRBM data,
$0.22\pm0.02$, well agrees with our KONUS measure (see Fig.~\ref{f:970228_var_bintime}),
but does not with R01 KONUS one.
The measure reported by R01 was derived from the three energy channels; this might
partially explain the difference. However, we notice that our KONUS measure
is 2.2-$\sigma$ apart from the R01 value of $0.08$. We are led to think of
two potential sources of discrepancy between our measure and R01's. First,
the overall time interval containing the GRB might be different; second, the
extrinsic scatter that R01 find on the global measure of $V_{f=0.45}$ is a little
underestimated with respect to what we find comparing a single KONUS channel with
the R01 global measure. We address the reader to the R01 paper for a definition of
the extrinsic scatter of variability due to the different energy channels derived
for each GRB.
Since we neither have the same KONUS data as R01, nor we know the overall time interval
adopted by R01, we cannot establish conclusively the reason for the discrepancy for
this GRB. However, concerning the first possibility, we tentatively adopted other time
intervals trying to match the variability measure reported by R01.
We find a variability measure of $0.08\pm0.03$ for a time interval including the first
sharp pulse and lasting about 40~s until the first pulse following a quiescent
interval from the very first pulse. It must be pointed out that our true measure
was performed on a 80-s long interval, since there is evidence for emission.
This could be a hint for the possible explanation of the discrepancy.

\subsubsection{GRB~991216}
For this GRB we adopted the measure obtained with GRBM data and we already discussed
the reasons in Sec.~\ref{ss:bintime}. Here we try to reproduce the R01 results using
the same BATSE data and then compare our variability measures on each energy channel with
the merged value derived by R01.
%
%Figure 3b
%
\begin{figure}
\includegraphics[width=9cm]{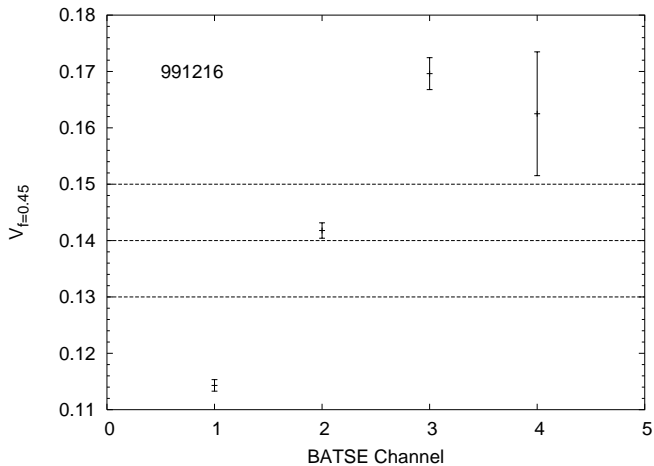}
\caption{Variability of 991216 as derived with BATSE data as a function of the
energy channel. Dashed lines show the merged value $\pm1~\sigma$ obtained by R01.}
\label{f:991216_batse_channels}
\end{figure}
In Figure~\ref{f:991216_batse_channels} we show the variability as a function of
the BATSE energy channel and compare them with the merged value $\pm1~\sigma$ reported
by R01. We remind that when comparing with GRBM results, we just considered channel 3.
In particular for this GRB, we know from previous discussion that the value obtained
with BATSE channel 3 appears to be underestimated with respect to the GRBM result
(see Fig.~\ref{f:991216_var_bintime}).

We have a perfect match within errors between our set of four smoothing time-scale
values and those obtained by R01. Therefore we are led to conclude that our variability
measures should match consequently. On this basis, from Fig.~\ref{f:991216_batse_channels}
we notice that the extrinsic scatter by R01, whose 1-$\sigma$ region is displayed through
dashed lines, seems to be little underestimated. In fact, the channel 1 measure is
2.6-$\sigma$ below and the channel 3 is 3-$\sigma$ above. 
In addition to this, we remind that for this particular GRB exhibiting sharp pulses
we know from GRBM data that a time binning of 64~ms is too coarse
(see Fig.~\ref{f:991216_var_bintime} and discussion in Sec.~\ref{ss:bintime}).
We therefore conclude that the effect of a higher scatter of variability at
different energy channels than that estimated by R01, combined with the fact the for
this GRB a binning time of 64~ms seems inadequate, account for the discrepancy between
our measure of variability for 991216 and that published by R01.

\subsubsection{GRB~000131}
\begin{figure}
\includegraphics[width=9cm]{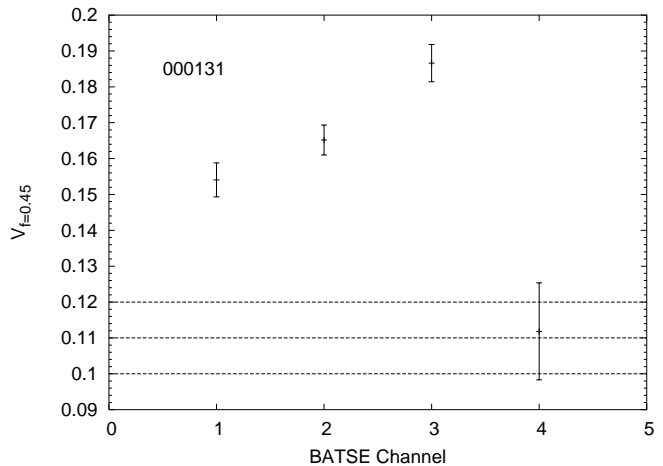}
\caption{Variability of 000131 as derived with BATSE data as a function of the
energy channel. Dashed lines show the merged value $\pm1~\sigma$ obtained by R01
with KONUS data.}
\label{f:000131_batse_channels}
\end{figure}

For this GRB we made use of BATSE data while R01 used KONUS data. Unfortunately
the public KONUS data of this GRB do not cover the whole profile, so we are
bound to use BATSE data alone and compare our variability measures with the
R01 value. Figure~\ref{f:000131_batse_channels} displays the variability as a
function of the BATSE energy channel and dashed lines show the R01 estimate.
%
%Figure 3c
%
The reasons of the discrepancy, which is apparent from Fig.~\ref{f:000131_batse_channels},
are due to the different smoothing time-scales: by comparing our set of four
values with the three ones corresponding to the three lower KONUS channels (the light
curve of channel 4 cannot be used according to R01), our values are systematically
greater than R01's. If we adopt the same time-scales obtained by R01 we get variability
measures which are consistent within the scatter with the R01 value. This conclusively
proves that the discrepancy for this GRB must be ascribed to the different measures
of the time-scales.
Concerning the origin of this discrepancy in the time-scales evaluation, we do not
find any apparent bias that could have affected the calculations using BATSE profiles.

\subsubsection{GRB~050315: a BAT/{\it Swift} GRB}
\begin{figure}
\includegraphics[width=9cm]{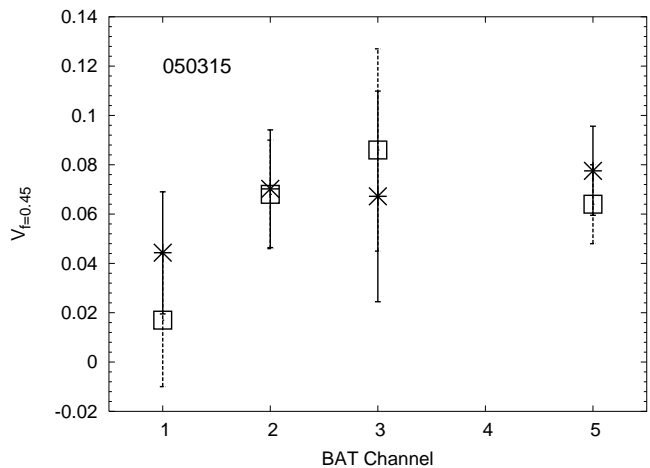}
\caption{Variability of 050315 as derived with BAT/{\it Swift} data as a function of the
energy channel. Asterisks and solid lines show our measures, while squares and dashed
lines show the measures by \citet{Donaghy05}.}
\label{f:050315_bat_channels}
\end{figure}

The variability measured for this BAT/{\it Swift} GRB is consistent with that
published by \citet{Donaghy05}. Here we want to show the consistency of the variability
measure with BAT/{\it Swift} data as well as its dependence on the the different
BAT energy channels. Figure~\ref{f:050315_bat_channels} shows the
variability as a function of the BAT channels obtained by us (asterisks and
solid lines) and by \citet{Donaghy05} (squares and dashed lines).
%
%Figure 3d
%
The energy bands of the three BAT channels considered are the following:
15--25~keV, 25--50~keV and 50--100~keV, respectively.
Channel 5 in Fig.~\ref{f:050315_bat_channels} corresponds to the integrated
band 15--100~keV. These energy channels have been chosen in order to match
those used by \citet{Donaghy05}.
Clearly the two sets of variability measures are consistent within errors
for each single BAT channels.

\subsubsection{Variability/Peak Luminosity}
\begin{figure*}
\centerline{\includegraphics[width=16cm]{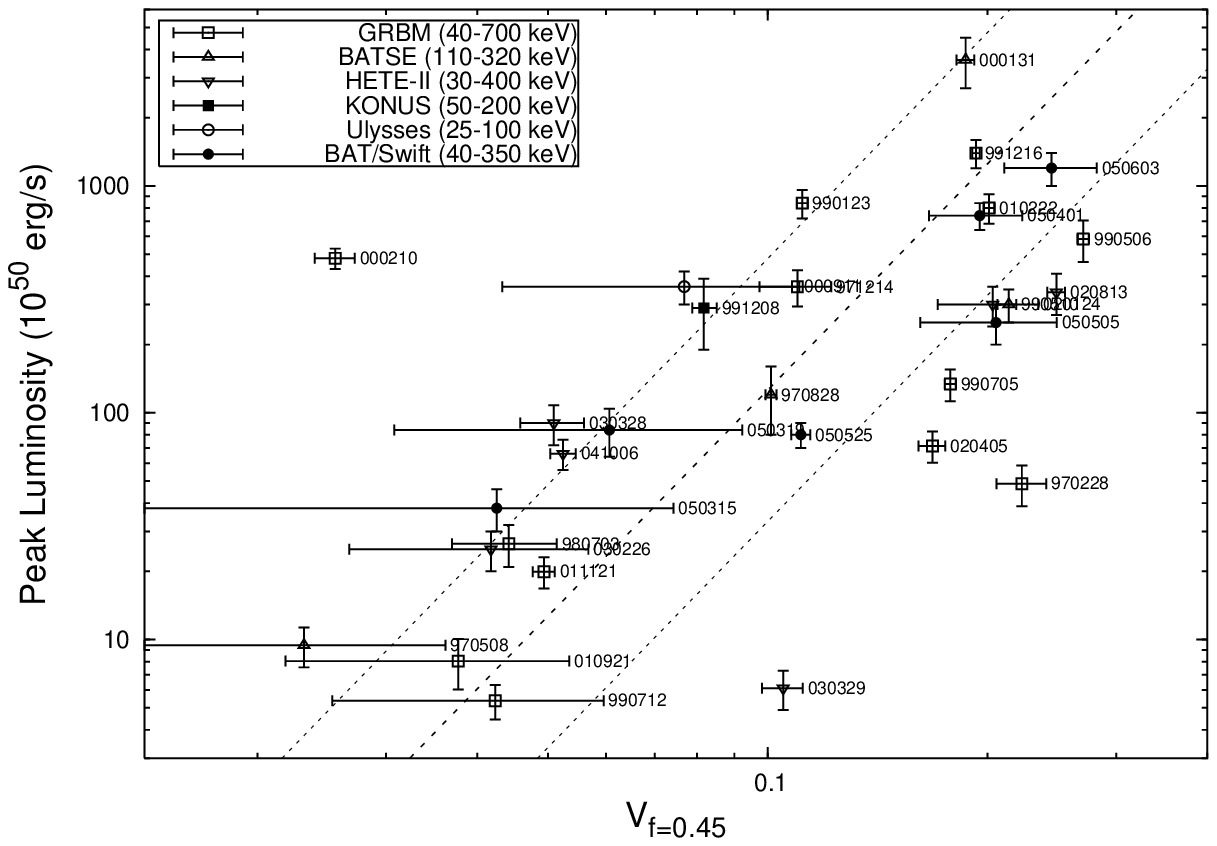}}
\caption{$V_{f=0.45}$ vs. Peak Luminosity for GRBs with known redshift. 
Dashed lines mark the best-fitting power-law relationship found by R01 (central
line) and  $\pm 1\sigma$ widths.}
\label{f:reichart_lum}
\end{figure*}

Figure~\ref{f:reichart_lum} and Table~\ref{tab:total_z} show the $V_{f=0.45}$ 
vs. Peak Luminosity for the entire sample of 32 GRBs with known redshift.
Dashed lines show the best-fitting power-law relationship found by R01 along with the
$\pm 1\sigma$ width, according to which $L \propto V_{{\rm R01}}^{m}$,
where $m=3.3_{-0.9}^{+1.1}$.
%
% Figure 4
%
Apparently, from Fig.~\ref{f:reichart_lum}, the correlation between the GRB variability
and the peak luminosity is confirmed, as also demonstrated by
the correlation coefficients and their statistical significance. 
The results are given in Table~\ref{tab:corr_coeff_var_reichart}, where we
report the values of both the Pearson linear correlation coefficient $r$, 
and the non-parametric correlation coefficients $r_s$ (Spearman rank-order coefficient) 
and $\tau$ (Kendall coefficient) \citep{Press93}, along with the corresponding
correlation statistical significance.
The same correlation coefficients have been evaluated also taking into account error
bars on both $V_{f=0.45}$ and $L$ through simulations (reported among brackets in
Table~\ref{tab:corr_coeff_var_reichart}). We scattered each point along with its error
bars assuming a Gaussian probability distribution in both dimensions and then we
calculated the mode for each coefficient distribution.

%
% Table 2
%
\begin{table}
\centering
% \begin{minipage}{150mm}
  \caption{Correlation Coefficients for GRBs with known redshift.
    We also report among brackets the mode values obtained from simulations.}
  \label{tab:corr_coeff_var_reichart}
  \begin{tabular}{lll}
\hline
Kind & Coefficient & Probability \\
\hline
Pearson's $r$    & 0.514 (0.412) & 0.0026 (0.019) \\
Spearman's $r_s$ & 0.625 (0.612) & 0.0001 (0.0002) \\
Kendall's $\tau$ & 0.446 (0.436) & 0.0003 (0.0005) \\
\hline
\end{tabular}
%\end{minipage}
\end{table}

\begin{figure}
\centerline{\includegraphics[width=9cm]{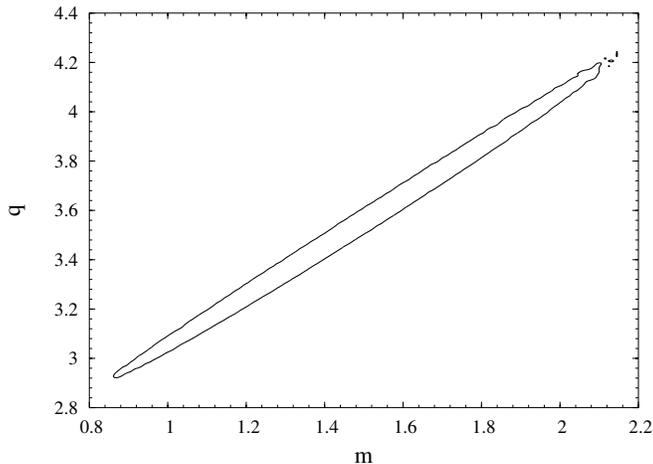}}
\caption{Contour plot of the 1$\sigma$ region of the two best-fitting parameters
$m$ and $q$ (least-square fit) in the case of $V_{f=0.45}$-$L$.}
\label{f:reichart_contour_vartf}
\end{figure}

However the high spread of the data points, clustered in two main regions of
the parameter space, shows that not only the best-fitting power-law parameters 
obtained by R01 are in disagreement with the our results but also that a 
power-law model gives a bad description of the data.
Indeed by fitting the data with a power-law model:
\begin{equation}
\log{L_{50}} \ = \ m\,\log{V_{f=0.45}} \ + \ q
\label{eq:bestfit_reichart}
\end{equation}
where the peak luminosity is $L=L_{50}\times 10^{50}$erg~cm$^{-2}$s$^{-1}$,
independently of the method used for the fit (usual least-square fit 
or minimization of absolute deviations, see \citet{Press93}),
we find unsatisfactory results ($\chi^2>1000$, 30 dof, in either cases).
Just to compare our best-fitting power-law model results with those obtained by R01,
in Table~\ref{tab:powerlaw_coeff_var_reichart} we report the best-fitting parameters
of the power-law for the two fit methods above mentioned.
In Fig.~\ref{f:reichart_contour_vartf} we report the  $1\sigma$ contour plot
of the best-fitting parameters $m$ and $q$. As can be seen, the two parameters are
highly correlated.

%
% Fig. 5
%
%
% Table 3
%
\begin{table}
 \centering
% \begin{minipage}{150mm}
  \caption{best-fitting power-law parameters of the $L$ vs. $V_{f=0.45}$ correlation
  for GRBs with known redshift.}
  \label{tab:powerlaw_coeff_var_reichart}
  \begin{tabular}{lrrr}
\hline
Method 		             & $m$       	     & $q$  &  $\chi^2/{\rm dof}$\\
\hline

%Least Sq.(single fit)       & $0.90 \pm 0.06$	& $3.16 \pm 0.06$ &   xxx/yyy\\
Least-square fit   & $1.30_{-0.44}^{+0.84}$ & $3.36_{-0.43}^{+0.89}$ & 1167/30 \\
Least-absolute-deviation fit & $1.16_{-0.17}^{+0.53}$ & $3.32_{-0.15}^{+0.49}$ & 1145/30\\
\hline
\end{tabular}
%\end{minipage}
\end{table}

We also evaluated the statistical uncertainty in the $\log{L_{50}}$ as a 
function of $V_{f=0.45}$, taking into account the correlation between the two parameters.
In Fig.~\ref{f:reichart_lum} the point corresponding to GRB 980425 is out of
the plot window to avoid scale compression, but its variability is
affected by a large uncertainty ($0.049_{-0.048}^{+0.048}$)
(see Table~\ref{tab:total_z}).

\subsection{Luminosity correction for GRB beaming}
\label{s:beaming}
\begin{figure*}
\includegraphics[width=16cm]{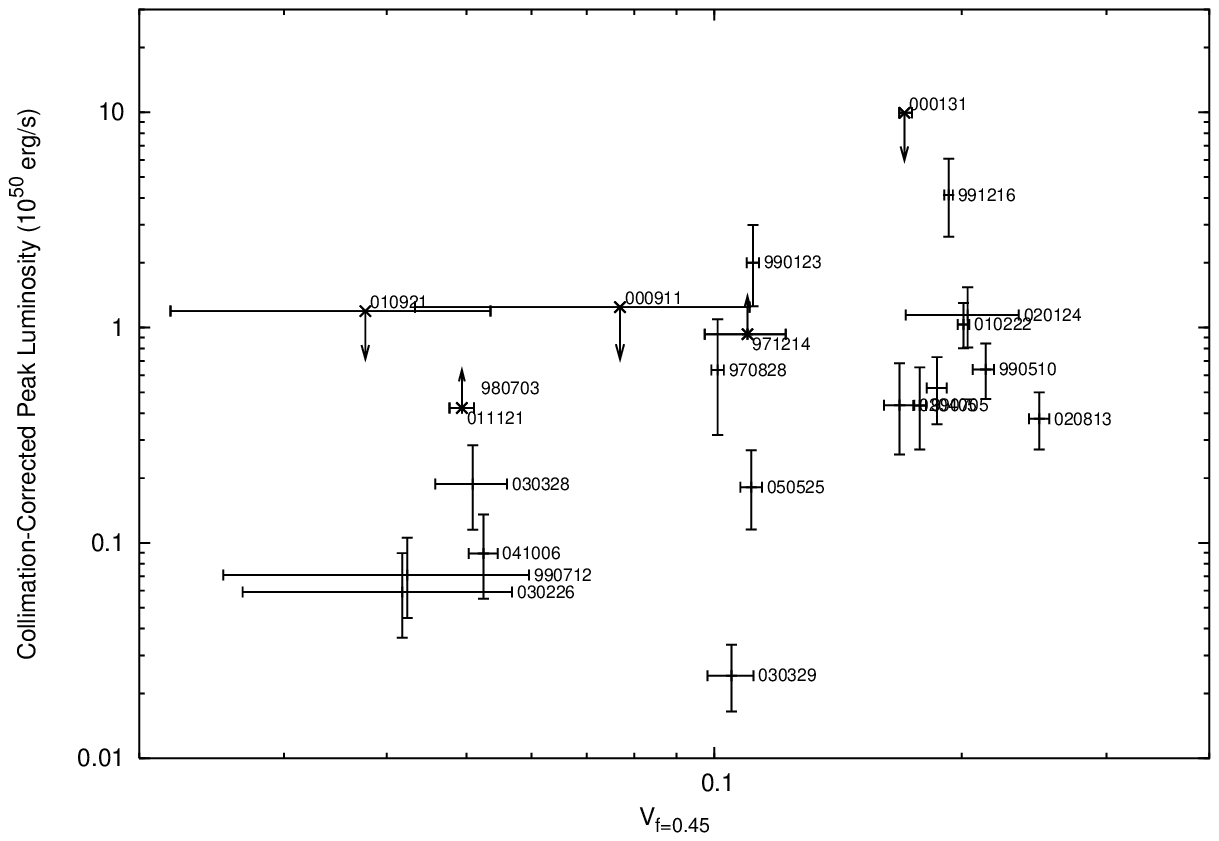}
\includegraphics[width=16cm]{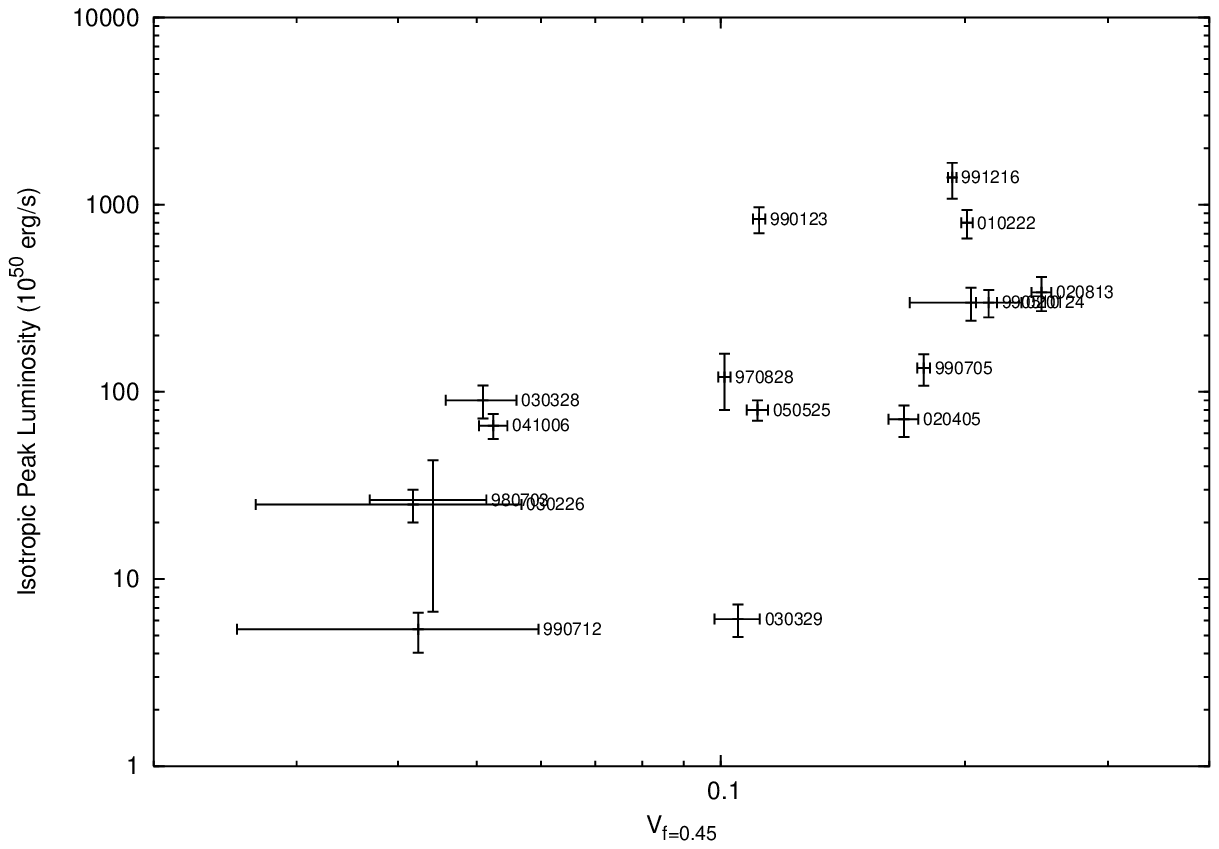}
\caption{{\em Top Panel}: Beaming-corrected rest-frame Peak Luminosity $L_{{\rm p},\gamma}$ vs.
Variability for a subset of 16 GRBs with known redshift and beaming angle \citep{Ghirlanda04}.
Also shown are two lower limits (971214 and 011121) and three upper limits (000131, 000911
and 010921).
{\em Bottom Panel}: $L_{\rm p}$ isotropic-equivalent Peak Luminosity vs. Variability for the
same 16 GRBs.}
\label{f:beaming}
\end{figure*}

For the GRBs with known redshift, we also investigated the correlation between
variability $V_{f=0.45}$ and peak luminosity after correcting the luminosity
values given in Table~\ref{tab:total_z} for the GRB beaming angles estimated
by \citet{Ghirlanda04}
\footnote{For a couple of them, i.e. 041006 and 050525, the values derived
by the same authors are taken from the following web site:
http://www.merate.mi.astro.it/$\sim$ghirla/deep/blink.htm}.
\citet{Ghirlanda04} demonstrated that after this correction,
the correlation between the peak energy $E_{\rm p}^{{\rm rest}}$ in the
source frame and $E_{{\rm rad},\gamma}$ (corrected for beaming) improved with a lower
spread of the data point around the best-fitting curve.
From our sample of GRBs with known redshift, we considered those for which
\citet{Ghirlanda04} provided the beaming angles: the resulting subset includes 16 GRBs.
In our case the result is shown in Fig.~\ref{f:beaming}. Unlike the findings by
\citet{Ghirlanda04} for the \citet{Amati02} relation, in the
present case the spread of the data points becomes larger when the energy released 
in the GRB is corrected for beaming, although the correlation remains significant
within 1\%. We computed the correlation coefficients for this
subset of GRBs in both cases: either assuming beaming-corrected $L_{{\rm p},\gamma}$ and isotropic
equivalent $L_{\rm p}$ peak luminosities vs. variability (see Table~\ref{tab:corr_coeff_beaming}).
%
% Figure 6
%
%

%
% Table 4
%
\begin{table}
 \centering
  \caption{Correlation Coefficients for 16 GRBs with known redshift and beaming angle:
$V$ vs. beaming-corrected $L_{{\rm p},\gamma}$ (first two columns) and $V$ vs.
isotropic-equivalent $L_{\rm p}$ (last two columns).}
  \label{tab:corr_coeff_beaming}
  \begin{tabular}{lcccc}
\hline
Kind & \multicolumn{2}{c}{$V$ vs. $L_{{\rm p},\gamma}$} & \multicolumn{2}{c}{$V$ vs. $L_{\rm p}$}\\
     & Coeff. & Prob. & Coeff. & Prob.\\
\hline
Pearson's $r$    & 0.664 & 0.005 & 0.688 & 0.003\\
Spearman's $r_s$ & 0.653 & 0.006 & 0.773 & 0.0005\\
Kendall's $\tau$ & 0.467 & 0.012 & 0.577 & 0.002\\
\hline
\end{tabular}
\end{table}
As reported in Table~\ref{tab:corr_coeff_beaming} and clearly shown by Fig.~\ref{f:beaming},
in the case of the isotropic-equivalent peak luminosity the spread is smaller than
in the case when the correction for beaming is applied.

%%%%%%%%%%%%%%%%%%%%%%%%%%%%%%%%%%%%%%%%%%%%%%%%%%%%%%%%%
\section{Discussion}
\label{s:disc}
%%%%%%%%%%%%%%%%%%%%%%%%%%%%%%%%%%%%%%%%%%%%%%%%%%%%%%%%%
The results found are puzzling. We confirm the peak luminosity
vs. variability correlation found by R01 when we use their sample of
GRBs, but we find a much larger spread of the data points when a
larger sample (32 events) of GRBs is used. In this case the
correlation between $V_{f=0.45}$ and $L$ is confirmed (significance
$\le$3$\times10^{-4}$ according to non-parametric tests),
but the data points are spread out in only two
regions of the parameter space, with a bad description of the data
points ($\chi^2 > 1000$, 30 dof) with a power-law function, which was the best-fitting
function found by R01. If, in spite of that, this function is used as 
fit model, the power-law index derived from our data ($m = 1.3_{-0.4}^{+0.8}$) is
much lower than that found by R01 ($m = 3.3_{-0.9}^{+1.1}$) and inconsistent with
it. 
The correlation becomes less significant (see the comparison between the two sets
of correlation coefficients in Table~\ref{tab:corr_coeff_beaming}) when we correct
the isotropic-equivalent peak luminosity for the GRB beaming, in contrast with the
result found by \citet{Ghirlanda04} who find a lower spread of the \citet{Amati02}
relationship when they perform this correction.

%%%%%%%%%%%%%%%%%%%%%%%%%%%%%%%%%%%%%%%%%%%%%%%%%%%%%%%%%
\section{Conclusions}
%%%%%%%%%%%%%%%%%%%%%%%%%%%%%%%%%%%%%%%%%%%%%%%%%%%%%%%%%
We have tested the correlation found by R01 between peak luminosity and
time variability following the same method used by R01 with a larger sample of
GRBs. For 32 GRBs with known redshift we confirm the existence of 
a correlation between the measure of time variability defined by R01
and the isotropic-equivalent peak luminosity. However we find a much higher
spread of the data points in the parameter space, with the consequence that
the correlation cannot be described by a power-law function as found by R01.
If, in spite of that, we fit the data with this function we find that
the power-law index ($1.3^{+0.8}_{-0.4}$) is much lower and
inconsistent with that found by R01 ($3.3^{+1.1}_{-0.9}$).
If we correct the peak luminosity for the GRB beaming, the correlation
is less significant.

%%%%%%%%%%%%%%%%%%%%%%%%%%%%%%%%%%%%%%%%%%%%%%%%%%%%%%%%%
\section*{Acknowledgments}
%%%%%%%%%%%%%%%%%%%%%%%%%%%%%%%%%%%%%%%%%%%%%%%%%%%%%%%%%
CG and AG acknowledge their Marie Curie Fellowships from the European Commission.
CG, FF, EM, FR and LA acknowledge support from
the Italian Space Agency and Ministry of University and
Scientific Research of Italy (PRIN 2003 on GRBs).
KH is grateful for {\it Ulysses} support under JPL contract 958056.
CGM acknowledges financial support from the Royal Society.
This research has also made use of data obtained from the {\it HETE2} science team via
the website http://space.mit.edu/HETE/Bursts/Data and BATSE, Konus/{\it WIND} and
BAT/{\it Swift} data obtained from the High-Energy Astrophysics Science Archive
Research Center (HEASARC), provided by NASA Goddard Space Flight Center.
%HETE is an international mission of the NASA Explorer program,
%run by the Massachusetts Institute of Technology.

%%%%%%%%%%%%%%%%%%%%%%%%%%%%%%%%%%%%%%%%%%%%%%%%
% BIBLIOGRAPHY
%%%%%%%%%%%%%%%%%%%%%%%%%%%%%%%%%%%%%%%%%%%%%%%%

\bsp

\label{lastpage}


\begin{thebibliography}{99}

\bibitem[\protect\citeauthoryear{Amati et~al.}{2000}]{Amati00}
Amati L. et~al., 2000, Sci, 290, 953
% GRB990705
%
\bibitem[\protect\citeauthoryear{Amati et~al.}{2002}]{Amati02}
Amati L. et~al., 2002, A\&A, 390, 81
%
\bibitem[\protect\citeauthoryear{Andersen et~al.}{2000}]{Andersen00}
Andersen M.I. et~al., 2000, A\&A, 364, L54
% GRB00131
%
%\bibitem[\protect\citeauthoryear{Antonelli et~al.}{2000}]{Antonelli00}
%Antonelli L.A. et~al., 2000, ApJ, 545, L39
%% GRB000214
%
\bibitem[\protect\citeauthoryear{Aptekar et~al.}{1995}]{Aptekar95}
Aptekar R.L. et~al., 1995, Space Sci. Rev., 71, 265
% Konus/WIND
%
\bibitem[\protect\citeauthoryear{Atteia}{2003}]{Atteia03a}
Atteia J.-L., 2003, A\&A, 407, L1
%
\bibitem[\protect\citeauthoryear{Atteia et~al.}{2003}]{Atteia03b}
Atteia J.-L. et~al., 2003, AIP Conf. Ser. Vol. 662, A Workshop Celebrating the First
Year of the HETE Mission, p. 17
% FREGATE/HETE-II
%
\bibitem[\protect\citeauthoryear{Bagoly et~al.}{2003}]{Bagoly03}
Bagoly Z., Csabai I., M\'esz\'aros A., M\'esz\'aros P., Horv\'ath I.,
Bal\'azs L.G., Vavrek R., 2003, A\&A, 398, 919
%
\bibitem[\protect\citeauthoryear{Band et~al.}{1993}]{Band93}
Band D. et~al., 1993, ApJ, 413, 281
%
\bibitem[\protect\citeauthoryear{Berger et~al.}{2005}]{Berger05_a}
Berger E., Bradley Cenko S., Steidel C., Reddy N., Fox D.B., 2005,
GCN 3368
% GRB050505
%
\bibitem[\protect\citeauthoryear{Berger \& Becker}{2005}]{Berger05_b}
Berger E. \& Becker G., 2005, GCN 3520
% GRB050505
%
\bibitem[\protect\citeauthoryear{Beuermann et~al.}{1999}]{Beuermann99}
Beuermann K. et~al., 1999, A\&A, 352, L26
% GRB990510
%
\bibitem[\protect\citeauthoryear{Bloom et~al.}{2003}]{Bloom03}
Bloom J.S., Berger E., Kulkarni S.R., Djorgovski S.G., Frail D.A.,
2003, AJ, 125, 999
% GRB990506
%
\bibitem[\protect\citeauthoryear{Boella et~al.}{1997}]{Boella97}
Boella G., Butler R.C., Perola G.C., Piro L., Scarsi L., Bleeker, 1997, A\&AS,  122, 299
%
\bibitem[\protect\citeauthoryear{Costa et~al.}{1998}]{Costa98}
Costa E. et~al., 1998, Adv.Sp.Res., 22, 1129 
%
\bibitem[\protect\citeauthoryear{Dado, Dar \& De R\'ujula}{2002}]{Dado02}
Dado S., Dar A., De R\'ujula A., 2002, A\&A, 388, 1079
%
%\bibitem[\protect\citeauthoryear{De Pasquale et al.}{2003}]{DePasquale03}
%De Pasquale M. et~al., 2003, ApJ, 592, 1018
%% Comparative Study of the X-ray afterglow properties of optically bright
%% and dark GRBs
%
\bibitem[\protect\citeauthoryear{Djorgovski et al.}{1998}]{Djorgovski98}
Djorgovski S.G., Kulkarni S.R., Bloom J.S., Goodrich R., Frail D.A.,
Piro L., Palazzi E., 1998, ApJ, 508, L17
% GRB980703
%
\bibitem[\protect\citeauthoryear{Djorgovski et al.}{1999}]{Djorgovski99}
Djorgovski S.G., Kulkarni S.R., Bloom J.S., Frail D.A., 1999, GCN 289
% GRB970228
%
\bibitem[\protect\citeauthoryear{Djorgovski et al.}{2001a}]{Djorgovski01a}
Djorgovski S.G., Frail D.A., Kulkarni S.R., Bloom J.S., Odewahn S.C.,
Diercks A., 2001a, ApJ, 562, 654
% GRB970828
%
\bibitem[\protect\citeauthoryear{Djorgovski et al.}{2001b}]{Djorgovski01b}
Djorgovski S.G. et~al., 2001b, GCN 1108
% GRB010921
%
\bibitem[\protect\citeauthoryear{Dodonov et al.}{1999}]{Dodonov99}
Dodonov S.N., Afanasiev V.L., Sokolov V.V., Moiseev A.V.,
Castro-Tirado A.J., 1999, GCN 475
% GRB991208
%
\bibitem[\protect\citeauthoryear{Donaghy et al.}{2005}]{Donaghy05}
Donaghy T.Q. et~al., 2005, GCN 3128
% GRB050315: GCN on variability
%
\bibitem[\protect\citeauthoryear{Fenimore et~al.}{1995}]{Fenimore95}
Fenimore E.E., In~'t Zand J.J.M., Norris J.P., Bonnell J.T.,
\& Nemiroff R.J. 1995, ApJ, 448, L101
%
\bibitem[\protect\citeauthoryear{Fenimore \& Ramirez-Ruiz}{2000}]{Fenimore00}
Fenimore E.E., Ramirez-Ruiz E., 2000, preprint (astro-ph/0004176)
%
\bibitem[\protect\citeauthoryear{Feroci et~al.}{1997}]{Feroci97}
Feroci M. et~al., 1997, SPIE Conf. Ser. Vol. 3114, p. 186
%
\bibitem[\protect\citeauthoryear{Foley et~al.}{2005}]{Foley05}
Foley R.J., Chen H.-W., Bloom J., Prochaska J.X., 2005, GCN 3483
% GRB050525
%
\bibitem[\protect\citeauthoryear{Frontera \& Fuligni}{1978}]{Frontera78}
Frontera F., Fuligni F., 1978, Nucl. Instr. Meth., 157, 557
%
\bibitem[\protect\citeauthoryear{Frontera \& Fuligni}{1979}]{Frontera79}
Frontera F., Fuligni F., 1979, ApJ, 232, 590
%
\bibitem[\protect\citeauthoryear{Frontera et~al.}{1997}]{Frontera97}
Frontera F. et~al., 1997, A\&AS, 122, 357
%
\bibitem[\protect\citeauthoryear{Fugazza et~al.}{2004}]{Fugazza04}
Fugazza D. et~al., 2004, GCN 2782
% GRB041006
%
\bibitem[\protect\citeauthoryear{Fynbo et~al.}{2005a}]{Fynbo05a}
Fynbo J.P.U., Hjorth J., Jensen B.L., Jakobsson P., Moller P., 2005a, GCN 3136
% GRB050319
%
\bibitem[\protect\citeauthoryear{Fynbo et~al.}{2005b}]{Fynbo05b}
Fynbo J.P.U. et~al., 2005b, GCN 3176
% GRB050401
%
\bibitem[\protect\citeauthoryear{Galama et~al.}{1999}]{Galama99}
Galama T.J. et~al., 1999, GCN 388
% GRB990712
%
\bibitem[\protect\citeauthoryear{Garnavich et~al.}{2001}]{Garnavich01}
Garnavich P.M., Pahre M.A., Jha S., Calkins M., Stanek K.Z.,
McDowell J., Kilgard R., 2001, GCN 965
% GRB010222
%
\bibitem[\protect\citeauthoryear{Gehrels}{2004}]{Gehrels04}
Gehrels N. et~al., 2004, ApJ, 611, 1005
%
\bibitem[\protect\citeauthoryear{Ghirlanda et~al.}{2004}]{Ghirlanda04}
Ghirlanda G., Ghisellini G., Lazzati D., 2004, ApJ, 616, 331.
% The collimation-corrected GRB energies correlate with the peak energy
% of their nuFnu spectrum
%
\bibitem[\protect\citeauthoryear{Greiner et~al.}{2003a}]{Greiner03a}
Greiner J., G\"unther E., Klose S., Schwarz R., 2003a, GCN 1886
% GRB030226
%
\bibitem[\protect\citeauthoryear{Greiner et~al.}{2003b}]{Greiner03b}
Greiner J., Peimbert M., Estaban C., Kaufer A., Vreeswijk P., Smoke J.,
Klose S., Reimer O., 2003b, GCN 2020
% GRB030329
%
\bibitem[\protect\citeauthoryear{Heise et~al.}{2001}]{Heise01}
Heise J., in 't Zand J.J.M., Kippen R.M., Woods P.M., 2001, in
Costa E., Frontera F., Hjorth J., eds, Proc. 2nd Rome Workshop,
Gamma Ray Bursts in the Afterglow Era, Springer-Verlag, Berlin, p. 16
%
\bibitem[\protect\citeauthoryear{Hjorth et~al.}{2003}]{Hjorth03}
Hjorth J. et~al., 2003, ApJ, 597, 699
% GRB020124
%
\bibitem[\protect\citeauthoryear{Hurley et~al.}{1992}]{Hurley92}
Hurley K. et~al., 1992, A\&AS, 92(2), 401
% Ulysses
%
\bibitem[\protect\citeauthoryear{Infante et~al.}{2001}]{Infante01}
Infante L., Garnavich P.M., Stanek K.Z., Wyrzykowski L., 2001, GCN 1152
% GRB011121
%
\bibitem[\protect\citeauthoryear{Ioka \& Nakamura}{2001}]{Ioka01}
Ioka K., Nakamura T., 2001, ApJ, 554, L163
%
\bibitem[\protect\citeauthoryear{Jager et~al.}{1997}]{Jager97}
Jager R. et~al., 1997, A\&AS, 125, 557
%
%\bibitem[\protect\citeauthoryear{Jakobsson et~al.}{2004}]{Jakobsson04}
%Jakobsson P., Hjorth J., Fynbo J.P.U., Watson D., Petersen K.,
%Bj\"ornsson G., Gorosabel J., 2004, ApJ, 617, L21
%
\bibitem[\protect\citeauthoryear{Kelson \& Berger}{2005}]{Kelson05}
Kelson D. \& Berger E., 2005, GCN 3101
% GRB050315
%
\bibitem[\protect\citeauthoryear{Kobayashi et~al.}{2002}]{Kobayashi02}
Kobayashi S., Ryde F., MacFadyen A., 2002, ApJ, 577, 302
% 
%
\bibitem[\protect\citeauthoryear{Kulkarni et~al.}{1998}]{Kulkarni98}
Kulkarni S.R. et~al., 1998, Nature, 393, 35
% GRB971214
%
\bibitem[\protect\citeauthoryear{Kulkarni et~al.}{1999}]{Kulkarni99}
Kulkarni S.R. et~al., 1999, Nature, 398, 389
% GRB990123
%
%\bibitem[\protect\citeauthoryear{Lazzati}{2002}]{Lazzati02}
%Lazzati D., 2002, MNRAS, 337, 1426
%% The role of photon scattering in shaping the light curves and spectra of GRBs
%
\bibitem[\protect\citeauthoryear{Le Floc'h et~al.}{2002}]{Lefloch02}
Le Floc'h E. et~al., 2002, ApJ, 581, L81
% GRB990705
%
\bibitem[\protect\citeauthoryear{Libert}{1978}]{Libert78}
Libert J., 1978, Nucl. Instr. Meth. 151(3), 555
% Statistique de comptage avec distribution uniforme de temps mort
%
\bibitem[\protect\citeauthoryear{Martini et~al.}{2003}]{Martini03}
Martini P., Garnavich P., Stanek K.Z., 2003, GCN 1980
% GRB030328
%
\bibitem[\protect\citeauthoryear{Masetti et~al.}{2002}]{Masetti02}
Masetti N., Palazzi E., Pian E., Hjorth J., Castro-Tirado A.,
Boehnhardt H., Price P., 2002, GCN 1330
% GRB020405
%
\bibitem[\protect\citeauthoryear{M\'esz\'aros et~al.}{2002}]{Meszaros02}
M\'esz\'aros P., Ramirez-Ruiz E., Rees M.J., Zhang B., 2002, ApJ, 578, 812
%
\bibitem[\protect\citeauthoryear{Metzger et~al.}{1997}]{Metzger97}
Metzger M.R., Djorgovski S.G., Kulkarni S.R., Steidel C.C., Adelberger K.L.,
Frail D.A., Costa E., Frontera, F., 1997, Nature, 387, 878
% GRB970508
%
\bibitem[\protect\citeauthoryear{M\"uller}{1973}]{Mueller73}
M\"uller J.W., 1973, Nucl. Instr. Meth. 112(1), 47
% Dead-time problems
%
\bibitem[\protect\citeauthoryear{M\"uller}{1974}]{Mueller74}
M\"uller J.W., 1974, Nucl. Instr. Meth. 117(2), 401
% Some formulae for a dead-time-distorted Poisson process 
%
\bibitem[\protect\citeauthoryear{Norris et~al.}{2000}]{Norris00}
Norris J.P., Marani G.F., Bonnell J.T., 2000, ApJ, 534, 248
% Connection between energy-dependent lags and peak luminosity in GRBs
%
\bibitem[\protect\citeauthoryear{Paciesas et~al.}{1999}]{Paciesas99}
Paciesas W.S. et~al., 1999, ApJS, 122(2), 465
% BATSE
%%
%
\bibitem[\protect\citeauthoryear{Piran}{2004}]{Piran04}
Piran T., 2004, preprint (astro-ph/0405503)
%
\bibitem[\protect\citeauthoryear{Piro et~al.}{2002}]{Piro02}
Piro L. et~al., 2002, ApJ, 577, 680
% GRB000210
%
\bibitem[\protect\citeauthoryear{Plaga}{2001}]{Plaga01}
Plaga R., 2001, A\&A, 370, 351
% The cepheid-like relationship between variability and luminosity
% explained within the ``cannonball model'' of GRBs
%
\bibitem[\protect\citeauthoryear{Press et~al.}{1993}]{Press93}
Press W.H., Flannery B.P., Teukolsky S.A., Vetterling W.T., 1993,
Numerical Recipes in C, Cambridge University Press, 2nd edition 
%
\bibitem[\protect\citeauthoryear{Price et~al.}{2002a}]{Price02a}
Price P.A. et~al., 2002a, ApJ, 573, 85
% GRB000911
%
\bibitem[\protect\citeauthoryear{Price et~al.}{2002b}]{Price02b}
Price P.A., Bloom J.S., Goodrich R.W., Barth A.J., Cohen M.H.,
Fox D.H., 2002b, GCN, 1475
% GRB020813
%
\bibitem[\protect\citeauthoryear{Reichart et~al.}{2001}]{Reichart01}
Reichart D.E., Lamb D.Q., Fenimore E.E., Ramirez-Ruiz E., Cline T.L.,
Hurley K., 2001, ApJ, 552, 57 (R01)
%
\bibitem[\protect\citeauthoryear{Reichart et~al.}{2003}]{Reichart03}
Reichart D.E., Lamb D.Q., Kippen R.M., Heise J., in 't Zand J.J.M.,
Nysewander M., 2004, in Feroci M., Frontera F., Masetti N., Piro L.,
eds, ASP Conf. Series, Vol. 312, Third Rome Workshop on Gamma Ray Bursts
in the Afterglow Era, Astron. Soc. Pac., San Francisco, p. 126
% preprint (astro-ph/0312501)
%
\bibitem[\protect\citeauthoryear{Sakamoto et~al.}{2004}]{Sakamoto04}
Sakamoto T. et~al, 2004, preprint (astro-ph/0409128)
%
\bibitem[\protect\citeauthoryear{Salmonson \& Galama}{2002}]{Salmonson02}
Salmonson J.D., Galama T.J., 2002, ApJ, 569, 682
%
\bibitem[\protect\citeauthoryear{Schaefer et~al.}{2001}]{Schaefer01}
Schaefer B.E., Deng M., Band D.L., 2001, ApJ, 563, L123
%
%\bibitem[\protect\citeauthoryear{Schlegel et~al.}{1998}]{Schlegel98}
%Schlegel D.J., Finkbeiner D.P., Davis M., 1998, ApJ, 500, 525
%
\bibitem[\protect\citeauthoryear{Tinney et~al.}{1998}]{Tinney98}
Tinney C., Stathakis R., Cannon R., Galama T., 1998, IAU Circ., 6896
% GRB980425
%
\bibitem[\protect\citeauthoryear{van der Klis}{1989}]{Klis89}
van der Klis M., 1989, in \"Ogelman H., van der Heuvel E.P.J., eds,
NATO ASI Ser. C 262, Timing Neutron Stars, Kluwer, Dordrecht, p. 27
%
\bibitem[\protect\citeauthoryear{Vreeswijk et~al.}{1999}]{Vreeswijk99}
Vreeswijk P.M. et~al., 1999, GCN 496.
% GRB991216
%

\end{thebibliography}
\end{document}